# Chemical Vapor Deposition Growth and Characterization of ReSe$_2$

J. Onasanya[b),1], Mourad Benamara[2], Kanagaraj Moorthi[2], H. O. H. Churchill[1,2,3], Bothina Hamad[a), 1,3], M. O. Manasreh[1,4]

[1]Materials Science and Engineering, University of Arkansas, Fayetteville, AR 72701, USA

[2]Institute for Nanoscale Science and Engineering, University of Arkansas, Fayetteville, AR 72701

[3]Department of Physics, University of Arkansas, Fayetteville, AR 72701, USA

[4]Department of Electrical Engineering and Computer Science, University of Arkansas, Fayetteville, AR 72701, USA

## Abstract

Two-dimensional (2D) flakes of ReSe$_2$ structure were grown by chemical vapor deposition and investigated at room temperature using Raman, photoluminescence, and absorption spectroscopies. The Raman spectra revealed eighteen phonon modes in the range of 100-300 cm$^{-1}$ that were found in good agreement with the density functional theory (DFT) calculations. The thickness profiles of the ReSe$_2$ flakes are in the range of 5-50 nm. The ReSe$_2$ crystal structure and morphology were investigated using XRD, atomic force microscopy and scanning electron microscopy. The energy dispersion spectroscopy confirmed the 1:2 elemental composition. The absorption spectra were obtained for ReSe$_2$ flakes and found to exhibit excitonic peaks in the spectral region of 885 - 942 nm (1.401 – 1.316 eV). These peaks are used to define the band gap of the material. The DFT calculations predicted an indirect bandgap of 0.88 eV for the bulk structure, while a direct bandgap of 1.26 eV was predicted for the monolayer.

*Keywords:* ReSe$_2$, absorbance, DFT, phonons, Raman

[a)]Email of corresponding authors: bothinah@uark.edu

[b)]Email of corresponding authors: onasanya@uark.edu

1. **Introduction**

Two-dimensional (2D) chalcogenide materials have attracted considerable interest due to their strong in-plane covalent bonds that contrasted with the weak van der Waals interlayer bonds. This van der Waals properties allow easy mechanical exfoliation at the atomic level where electron confinement plays a major role [1-3]. This crystallographic nature leads to intriguing electrical and optoelectronic device characteristics, such as carrier mobility, exciton binding energy, phonon scattering, and thermal conductivity [4-10]. Among these 2D materials, there is an emerging class of low-symmetry materials with orthorhombic, monoclinic, or triclinic crystal structures with unique in-plane lattice arrangements that leads to in-plane orientation-dependent physical properties [11-14]. The distinct properties of these materials exhibit anisotropic electronic and optical properties that are promising for applications in polarization-based optical devices [15, 16]. Black phosphorus (BP) is the first known low-symmetry 2D semiconductor layered material with a puckered honeycomb arrangement. It possesses strong in-plane optical anisotropy due to its orthorhombic crystal structure. Like other well-known 2D layered materials, it exhibits excellent properties for its potential application in optoelectronics [17]. However, there are limitations to the broader use of BP due to its poor environmental stability leading to degradation due to oxidation under ambient conditions [18]. Therefore, there is a challenge in synthesizing high-quality and large-area flakes [19]. These highlighted limitations of BP have inspired the research of alternative low-symmetry materials with environment stability such as rhenium dichalcogenides, $ReX_2$ (X= S, Se). Rhenium dichalcogenides fall into the class of transition metal dichalcogenides that crystallize in a distorted triclinic structure with a $\bar{P}1$ space group [20]. This structure corresponds to a distorted $CdCl_2$ type configuration [21, 22] caused by the Re-Re zig-zag chains formed within each layer, which breaks the in-plane symmetry. These properties lead to anisotropic optical and electronic properties, tunable bandgap, and pronounced nonlinear optical responses [23, 24], which are ideal for applications in ultrafast photonics and high-performance photodetectors, while enabling new applications in polarization-sensitive technologies [25, 26].

Several methods have been reported for synthesis of $ReX_2$ (X= S, Se), such as mechanical exfoliation, chemical vapor transport, liquid phase exfoliation, hydrothermal and chemical vapor deposition (CVD) [27] . Among these methods, the CVD technique has shown a precise control of material growth parameters, which is an essential feature for producing high-quality 2D materials such as rhenium diselenide ($ReSe_2$) [28]. Li *et. al* reported the CVD growth high quality vertically aligned $ReSe_2$ nanosheets directly on conductive substrates resulting in effective growth at temperatures near 450 °C [25, 29]

In this article, we report on the structural, electronic, and optical properties of high purity $ReSe_2$ flakes synthesized by CVD technique. The morphology and elemental composition were analyzed using scanning electron microscopy (SEM), atomic force microscopy (AFM) and energy dispersive spectroscopy (EDS). The obtained images of the synthesized flakes show a triclinic crystal symmetry. The experimentally observed Raman spectra, using μ-Raman spectroscopy, revealed 18 phonon modes that are in good agreement with the DFT calculated results. Optical absorption measurements showed distinct excitonic peaks, and X-ray diffraction (XRD) analysis confirmed the quality of the $ReSe_2$ crystals with sharp diffraction peaks. The successful growth of $ReSe_2$ flakes is encouraging toward the fabrication of optoelectronic devices.

2. **Experiment**

Rhenium diselenide ($ReSe_2$) single crystal flakes with various thicknesses were synthesized using CVD growth method. A 100 mg of ammonium perrhenate ($NH_4ReO_4$) with 99.9% purity and 500 mg of selenium (Se) powdered precursors were used. These precursors were purchased from Strem Chemicals Inc. Argon and hydrogen mixed gas in a ratio of 4:1 was used as the carrier for the mass transport of thermally evaporated chemical species inside the growth chamber. The argon-hydrogen gas mix was used as a carrier with a flow rate of 30 sccm. The growth process was conducted in a horizontal tube furnace at a temperature of 700 °C. The furnace was initially purged with Ar gas to remove any residual oxygen, ensuring an inert atmosphere. The growth time was maintained for 30 minutes to facilitate the formation of ReSe$_2$ flakes.

A Horiba LabRAM HR spectrometer was used to obtain μ-photoluminescence (PL) and μ-Raman spectra. A HeNe laser of a wavelength of 632.8 nm was used to investigate Raman vibrational modes. The spectra range of the data collected is between 100 – 300 cm$^{-1}$. A Cobalt laser of a wavelength of 472.9 nm was used for the PL measurement spanning spectral range of 450 – 1100 nm. Angle-resolved polarized Raman spectroscopy was carried out to determine the in-plane anisotropy and crystallographic orientation of the sample. Polarization rotation was performed by rotating a half-wave plate through rotation angle varied from 0° to 360° in 10° increments. At each angle, Raman spectra were acquired under identical integration conditions and processed to extract peak intensities. The absorbance spectrum was obtained using Cary 500 UV-Visible-NIR by Agilent. Structural analysis of the synthesized ReSe$_2$ was conducted using an X-ray diffractometer (Rigaku Miniflex) equipped with Cu Kα radiation (λ = 1.5406 Å). The diffraction patterns collected were over a 2θ range of 10–60° to identify the crystalline phases and assess the quality of the deposited layers. A Bruker D3100 Nanoscope AFM, equipped with a tapping mode module, was utilized to investigate the surface morphology of the grown ReSe$_2$ flakes. The SEM images were obtained by using a FEI Nova Nanolab 200 fitted with a Bruker EDS detector.

3.  **Theory**

Theoretical investigations were performed using DFT computational method as implemented in Vienna *ab initio* simulation package (VASP) [30, 31]. The generalized gradient approximation (GGA) with the Perdew–Burke–Ernzerhof (PBE) is adopted for the exchange-correlation functionals [32]. The Van der Waals interactions were considered in these calculations using the optB86b functional [33]. A plane-wave cutoff energy of 550 eV was employed with energy convergence criterion set at $10^{-8}$ eV. The Brillouin zone was sampled using 9×9×8 and 9×9×1 Monkhorst–Pack K -point grids for the bulk and monolayer systems, respectively. The density functional perturbation theory (DFPT) technique [34] was used to obtain the phonon modes at the gamma point. The layered stacking of $ReSe_2$ structure is composed of rhenium atoms sandwiched between the selenium atoms as shown in Fig. 1 (a) and (b). This figure shows the distorted triclinic structure known as Peierls distortion, where the Re atoms form Re-Re chains with diamond shape.

## 4. Results and Discussion

The AFM details of $ReSe_2$ flakes grown on $SiO_2$/Si substrate with the height distribution is represented by Fig. 2(a) in a range of values from 5-50 nm indicated by the profile height scan. The surface topography as presented in Fig (2b) reveals the flakes morphology with varying thickness across the scanned area. The line profile corresponds to the thickness as depicted by the color markers green, blue, and red have the values 50, 10, and 5 nm, respectively. It is observed that the bright island spots are thicker in dimensions than the dim ones. Therefore, the variation in height is suggestive of multilayered $ReSe_2$ formation ranging from few to several layers in agreement with Hefeez, *et al*. [35].

The surface morphology of $ReSe_2$ flakes is obtained by SEM, which shows well-defined hexagonal-like structure with varying thickness as indicated in Fig. 3. The presence of layered structures as seen from the SEM image is typical for van der Waals 2D layered materials along the basal plane direction [36]. The stoichiometry of the atomic percentage of Re:Se is 1:2 as confirmed by the EDS as reported in Table I.

The X-ray diffraction (XRD) spectrum as shown in Fig. 4 has well defined peaks at diffraction angles 14.4°, 28.5°, 42.9°and 58.2°, which are readily indexed as (002), (004), (0 0 6), and (0 0 8) crystal planes of the layered $ReSe_2$, respectively. These observed peaks positions are characteristics of crystalline, layered $ReSe_2$ with distorted triclinic crystal symmetry, which matches the ICDD card No. 04-007-1113 [24, 37]. We note the observation of no other noticeable peaks except the well resolved XRD patterns of $ReSe_2$, emphasizing the quality of the growth.

The μ-Raman spectra of the grown $ReSe_2$ were taken at room temperature as shown in Fig. 5. This spectrum revealed the presence of the expected and well-defined 18 Raman modes. The experimentally observed Raman modes are consistent with the computational values using the DFPT method [22, 25] as shown in Table II. These observed Raman modes mostly possess $A_g$ symmetry which corresponds to the in-plane vibrational modes [38, 39]. The active Raman modes of vibrations within the short range (100 – 300 $cm^{-1}$) further indicate the low symmetry properties of $ReSe_2$ with an anisotropic behavior due to its triclinic crystal symmetry. This anisotropic behavior distinguishes its optical and electronics responses [39] from other transition-metal dichalcogenides (TMDs) with high symmetry such as $WS_2$ and $MoS_2$.

The polarization polar plot of three selected Raman modes (125.3, 162, and 177 $cm^{-1}$) from the experimental data were analyzed to understand the optical anisotropy of the grown $ReSe_2$. The resulting polar plot takes the form of a two-lobe shape characteristics of anisotropic low-symmetric 2D materials [40]. However, the lobes display unequal intensity maxima, indicating a deviation from ideal symmetric anisotropic scattering. This can be attributed to the surface roughness, morphology [41] and thickness variation [42] from the growth process as evidenced by the SEM image in Fig. 3. This premise resulted in altering the incident polarization and the scattered intensities thereby modifying the effective Raman tensor. To account for the asymmetric behavior and describe the angular dependence of the Raman signal, the scattering intensity is expressed through the Raman tensor ($R$) and the unit polarization vectors of the incident ($e_i$) and scattered ($e_s$) light [43].

$$I \propto |e_i R e_s|^2 \qquad (1)$$

Equation (2) was used to fit the polar plot where $\theta$ is polarization incident angle made with crystal which is oriented along the distorted Re-Re chain where the maximum intensities are observed

$$I(\theta) \propto v^2 + u^2 \cos^2 C + w^2 \sin^2\theta + 2v(u+w)\cos\theta\sin\theta \qquad (2)\ [22]$$

where $u, v,$ and $w$ are the elements for each vibrational mode given by $2x2$ matrix $R = \begin{pmatrix} u & v \\ v & w \end{pmatrix}$ and the polarization angle, $\theta$, is obtained from the $\tan 2\theta = \frac{2v}{u-w}$. From the polarization-dependent Raman analysis, the angular dependence of the selected Raman modes (125.3, 162.3, and 177 cm$^{-1}$) was fitted using equation (2). The results are shown in Fig. 6 where the black squares represent the measured angle-resolved Raman intensities, and the solid red curves correspond to the tensor-based fitted profiles. The vibrational modes at 125.3 and 162.3 cm$^{-1}$ exhibit their maximum intensities near 80° and 260°, whereas the 177 cm$^{-1}$ mode displays a slightly shifted pair of maxima at approximately 75° and 255°. The observed 180° periodicity in the different modes is characteristic of the two-fold angular symmetry expected for strongly anisotropic 2D materials [44, 45]. The variation in the polarization angles of maximum intensity among the different modes indicates mode-dependent Raman tensor orientations and reflects the underlying preferential alignment of the distorted Re–Re atomic chains. The results of the fitting tensor parameters and the extracted polarization directions for each mode are summarized in Table III.

The non-zero and non-equivalent values of the tensor fitted parameters ensure the angular dependency for each vibrational modes, this outcome confirms the anisotropic nature of ReSe$_2$.

The room-temperature absorption measurement of the grown ReSe$_2$ is presented in Fig. 7. The absorption spectra exhibit two prominent excitonic peaks located at 942 nm (red line) and 885 nm (blue line), corresponding to energy transition with approximate values of 1.32 eV and 1.40 eV, respectively. These excitonic features indicate a slight layer-dependent variation in the optical bandgap. The obtained bandgap values are consistent with the results reported by Huang *et al.*, who observed temperature-dependent excitonic transitions ranging from 1.36 to 1.42 eV in polarization-resolved absorption measurements between 25 K and 525 K [46, 47]. The close agreement between our room-temperature measurements and their temperature-dependent study further validates that excitonic transitions exist regardless of temperature variation. The slight blue shift of the higher-energy peak to 1.40 eV in our measurement can be attributed to the reduced layer thickness in the grown flakes. The bandgap showed slight variation with thickness, suggesting that there is a weak degree of quantum confinement at the nanoscale [23, 48, 49]. However, it is not pronounced when compared to other traditional TMDs, such as MoS$_2$, WS$_2$, or ReS$_2$ where studies have demonstrated a strong quantum confinement effect [45, 50, 51].

The electronic band structure combined with the partial density of states for both bulk and monolayer ReSe$_2$, obtained using DFT calculations, are represented by Fig. 8(a, b). The energy bandgap of the bulk is found to be indirect, having an energy value of 0.88 eV [20, 22] with the maximum of the valence band located at the Γ points, while its conduction band minimum is located between R and Γ points. However, the monolayer band structure shows a direct bandgap of 1.26 eV at the Γ high symmetry point [52]. Equally, the partial density of states calculation showed the orbitals contribution of each element, which perfectly describe the electronic properties in both the bulk and monolayer form. The conduction band is dominated by the d-orbital of the Re atoms while the valence band has almost equal contribution of both the d-orbit of Re and p-orbit of Se atoms. The DFT predicted bandgaps showed slight observable increase in bandgaps. This further established the weak interlayer dependence of the bandgap as reported in other DFT based studies for ReSe$_2$ [53].

The room temperature PL measurement as represented by the black line in Fig. 9 exhibits a broad spectrum of optical transitions with one main peak around 770 nm (1.60 eV) and two shoulders at 670 nm (1.85 eV) and 860 nm (1.44 eV). This spectrum is fitted with a Gaussian line shape (see a magenta line) to indicate the presence of three main peaks. Gaussian curves (green, blue, and red) were added to resolve the presence of the three peaks. These three peaks represent excitonic transitions within the flake under the assumption that the electrons were not undergoing transitions from the valence band to different minima in the conduction band. The transitions are rather generated from the same flake with nonuniform thickness such that the quantum confinement is not uniform within the flake. The three excitonic transitions correspond to the band gap values. The experimentally reported results of the band gaps are found to be higher than those predicted by DFT calculations as shown in Fig. 7. This can be attributed to the approximation method associated with DFT, which leads to bandgap underestimation. However, the observed high energy excitonic value of 1.85 eV (green curve) from Fig. 9 is consistent with the first-principles GW-Bethe-Salpeter equation approach prediction in the study of ReSe$_2$ by Zhong *et al*. [54], which is around 2.09 eV for suspended monolayer ReSe$_2$. The latter band gap value is higher than the experimental results due to depressed screening and reduced dimension thereby enhancing quantum confinement [17, 55-57].

## 5. Conclusion

Highly crystalline anisotropic layered ReSe$_2$ flakes with hexagonal structures were successfully grown via CVD. The Raman scattering spectrum reveals the presence of 18 phonon modes in the range of 100 - 300 cm$^{-1}$, which agrees with the computational predictions, indicating accurate phonon dispersion in the material. Density Functional Theory (DFT) calculations predicted an indirect bandgap of 0.88eV for the bulk and 1.26eV for the monolayer of ReSe$_2$. The absorption spectra of the grown ReSe$_2$ were found to be in the near-infrared region, between 885-942 nm, suggesting layer-dependent optical response. The energy bandgap values obtained from photoluminescence measurement further confirm the layer dependent nature of ReSe$_2$ with an optical energy emission between 1.44 -1.85 eV, this implies its application ranges from visible to near infrared region-based optoelectronics devices. Scanning electron microscopy and energy-dispersive X-ray spectroscopy confirm the quality and morphology of the material with the profile height between 5-50 nm measured by AFM.

**Credit authorship contribution statement**

Juwon Onasanya (J.O.): Writing – original draft, Software, Methodology, Investigation, Data curation. Mourad Benamara: Validation, Resources, Formal analysis, Data curation. Kanagaraj Moorthi: Validation, Resources, Formal analysis, Data curation. Bothina Hamad (B.H.): Validation, Supervision, Software, Methodology, Funding acquisition, Conceptualization. H. O. H. Churchill: Validation, Project administration, Funding acquisition, Conceptualization. M.O. Manasreh: Validation, Supervision, Project administration, Funding acquisition, Conceptualization

**Declaration of Competing Interest**

The authors declare that they have no known competing financial interests or personal relationships that could have appeared to influence the work reported in this paper.

**Acknowledgment**

**B.H.** thanks the support of the MonArk Quantum Foundry that is funded by the National Science Foundation Q-AMASE-i program under NSF Award No. DMR-1906383. In addition, **B.H.** declares that this material is based upon work supported by the National Science Foundation under Award No. DGE-2244274. DMR-1906383. **J.O.** declares that this material is based on research sponsored by AFRL under agreement number FA8750-24-1-1019. The U.S. Government is authorized to reproduce and distribute reprints for Governmental purposes notwithstanding any copyright notation thereon. The authors would like to thank Arkansas High Performance Computing Center (AHPCC), University of Arkansas, for the computing resources.

**Data availability**

Data will be made available on reasonable request.

**Figure Caption**

Fig. 1(a) Top view showing the zig-zig arrangement of rhenium atoms in ReSe$_2$ structure (b) Side view ReSe$_2$ structure with the Rhenium atoms sandwiched between the selenium atoms.

Fig. 2 (a) Height profile measurement of scanned flakes, the colors correspond to each identified spot on the AFM image. (b) AFM image showing the grown ReSe$_2$ flakes.

Fig. 3 SEM image of ReSe$_2$ flakes with hexagonal-like layered structure.

Fig. 4 X-ray diffraction pattern of ReSe$_2$ flakes with its characteristic peaks indexed at (002), (004), (006), and (008) showing preferential orientations along the c-axis typical for 2D layered ReSe$_2$.

Fig. 5 The micro-Raman spectrum of ReSe$_2$ at room temperature with 18 observable phonon modes in the range of 100 – 300 cm$^{-1}$. The observed vibrational modes are consistent for a typical triclinic ReSe$_2$ and are in good agreement with DFT calculations.

Fig. 6 Angle-resolved polarized Raman polar plot of the intensity of the vibrational mode at (a) 125.3 (b) 162.3 cm$^{-1}$ and (c) 176.1 cm$^{-1}$ as a function of the polarization angle. Black squares indicate experimental data points, and the solid red line represents the theoretical fit derived from the Raman tensor analysis. (d) Optical microscope image of the scanned ReSe$_2$ flake for polarization measurements.

Fig. 7 The room temperature absorbance spectra plot of ReSe$_2$ flakes with observed maximum wavelength peaks at 885 and 942 nm.

Fig. 8 (a) The density functional theory calculation of combined plot of density of states and energy bandgap for bulk ReSe$_2$ with energy value of 0.88 eV. (b) The density functional theory calculation of combined plot of density of states and energy bandgap for monolayer ReSe$_2$ with energy value of 1.26 eV.

Fig. 9 The guassian fit of room temperature PL measurement of CVD grown ReSe$_2$ flake. The color black represent the experimental curve while the other colors green, blue, and red are the corresponding guassian curves for the excitonic peaks. The peaks value for the fitted curve are 670 nm (1.85 eV), 770 nm (1.61 eV) and 860 nm (1.44 eV).

Table I. Energy-dispersive spectroscopy elemental composition of $ReSe_2$ flakes.

| Elements | Series | Concentration (Atomic Weight %) | Error |
|---|---|---|---|
| Re | L- series | 24.59 | 1.63 |
| Se | L- series | 45.07 | 1.76 |

Table II. Raman modes for experimentally measured and DFT calculated for ReSe$_2$ both in the range of 100 -300 cm$^{-1}$ with a total of 18 raman active peaks.

| Peak Numbers | Experiment (cm$^{-1}$) | Calculated (cm$^{-1}$) | Raman Mode |
|---|---|---|---|
| 1 | 116.3 | 112.7 | A$_g$ |
| 2 | 122.1 | 117.9 | A$_g$ |
| 3 | 125.3 | 122.8 | A$_g$ |
| 4 | 128.5 | 125.5 | A$_g$ |
| 5 | 162.3 | 159.4 | A$_g$ |
| 6 | 176.1 | 172.0 | A$_g$ |
| 7 | 180.7 | 177.1 | A$_g$ |
| 8 | 184.2 | 179.9 | A$_g$ |
| 9 | 195.3 | 192.0 | A$_g$ |
| 10 | 200.0 | 195.1 | A$_g$ |
| 11 | 213.6 | 208.2 | A$_g$ |
| 12 | 221.5 | 219.7 | A$_g$ |
| 13 | 236.4 | 236.6 | A$_g$ |
| 14 | 243.5 | 241.3 | A$_g$ |
| 15 | 251.1 | 249.8 | A$_g$ |
| 16 | 264.5 | 262.3 | A$_g$ |
| 17 | 287.5 | 285.1 | A$_g$ |
| 18 | 297.3 | 295.3 | A$_g$ |

Table III. Angle-resolved polarized Raman-tensor fit ($u$, $v$, $w$) and polarization angle ($\theta$)

| Peak (cm$^{-1}$) | u | v | w | $\theta_{max}$ | $\theta_{max} + 180°$ |
|---|---|---|---|---|---|
| 125.3 | 2.40 | 13.02 | 70.95 | 80° | 260° |
| 162.1 | 9.39497 | 5.1785 | 32.25 | 77.8° | 257.8° |
| 176.3 | 15.000 | 10.53908 | 49.99 | 74.5° | 254.5 |

Table IV. Summary of the energy bandgap for PL gaussian fits, DFT-calculated bandgap, and experimental absorption for ReSe$_2$.

| | PL (eV) | Absorption (Expt.) (eV) | DFT-Bandgap (eV) |
|---|---|---|---|
| ReSe$_2$ (Bulk) | | | 0.88 |
| ReSe$_2$ (Monolayer) | | | 1.26 |
| Thick flakes | | 1.32 | |
| Thin flakes | | 1.40 | |
| Fit peak 3 | 1.44 | | |
| Fit peak 2 | 1.60 | | |
| Fit peak 1 | 1.85 | | |

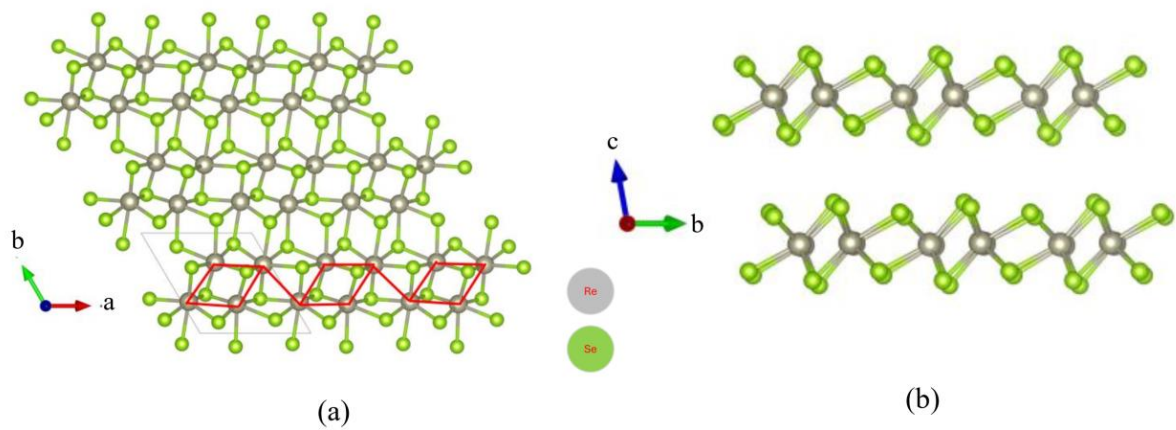

Fig. 1. Onasanya, *et al.*

(a) 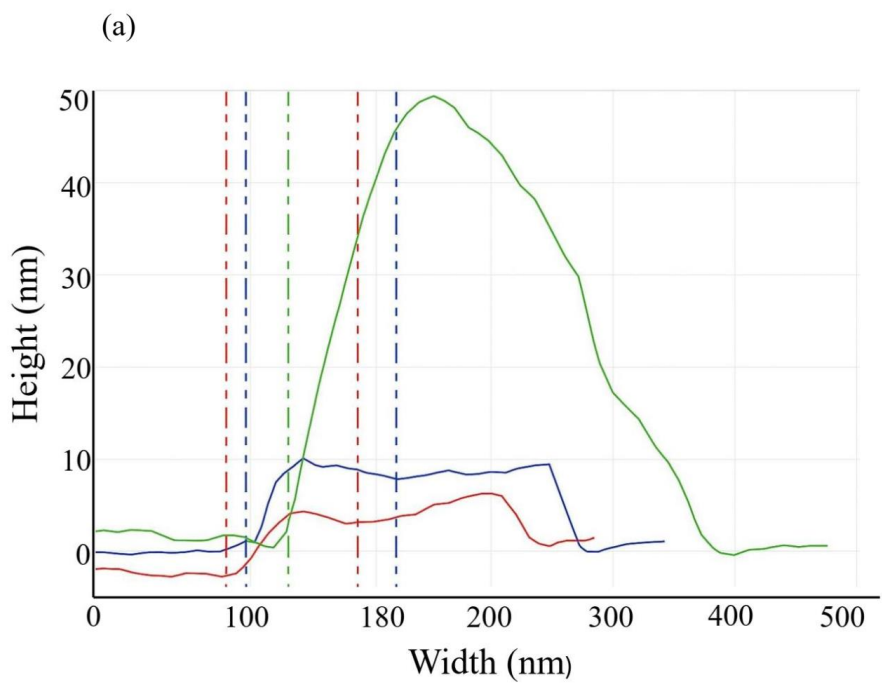

(b) 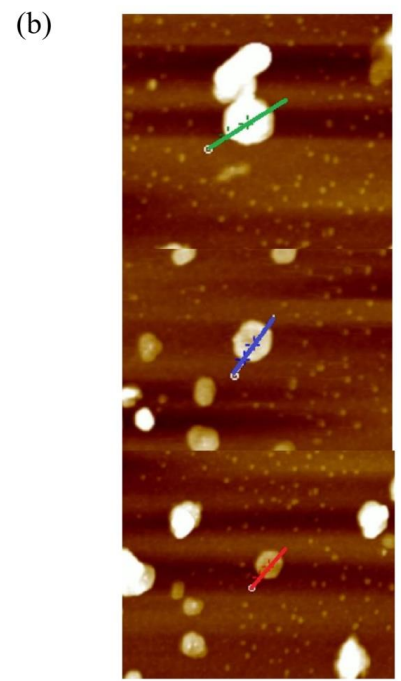

Fig. 2. Onasanya, *et al.*

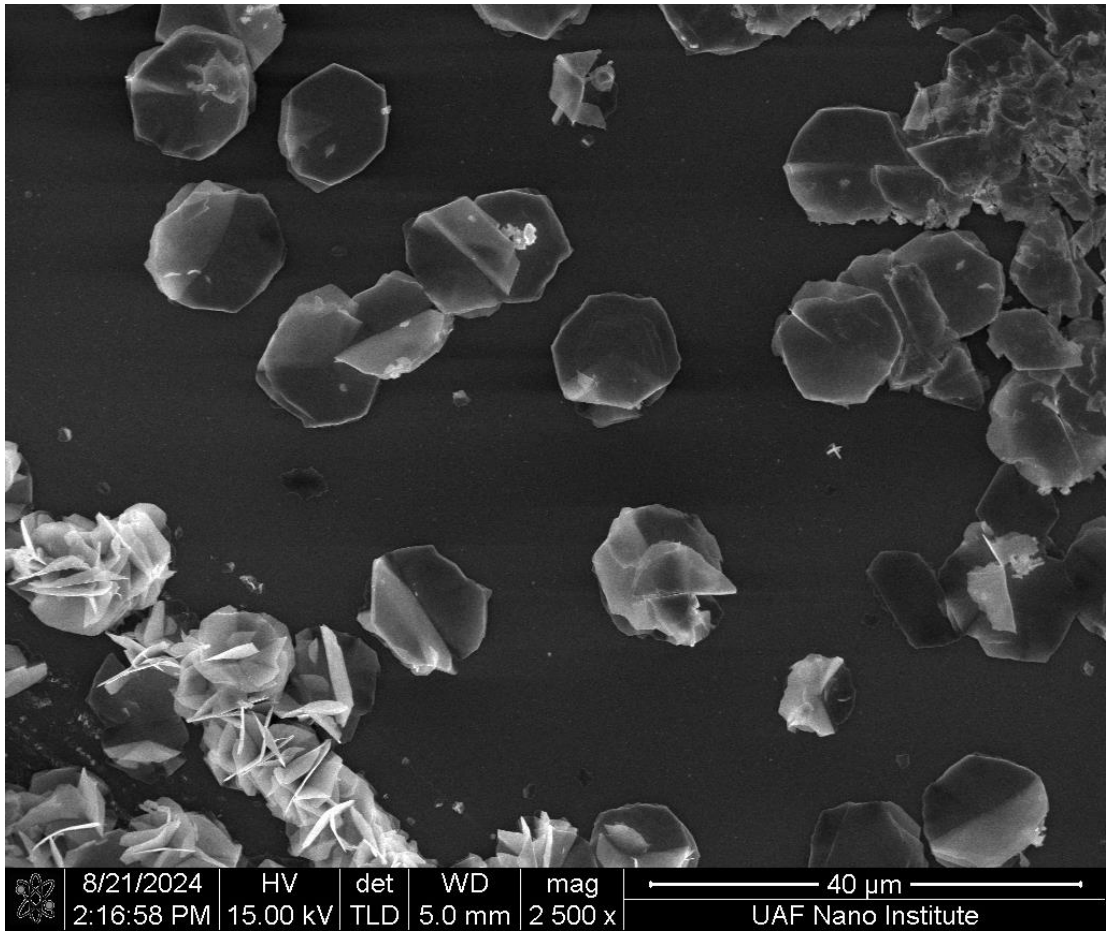

Fig. 3. Onasanya, *et al.*

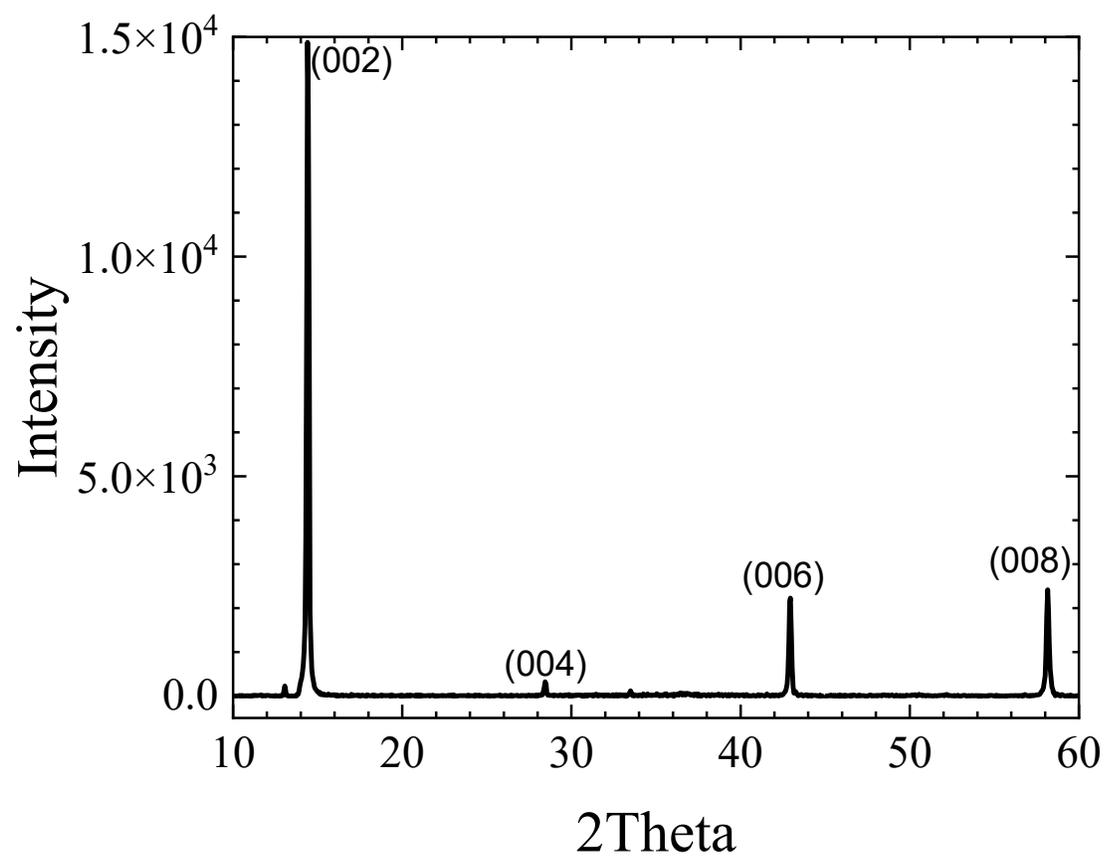

Fig. 4. Onasanya, *et al.*

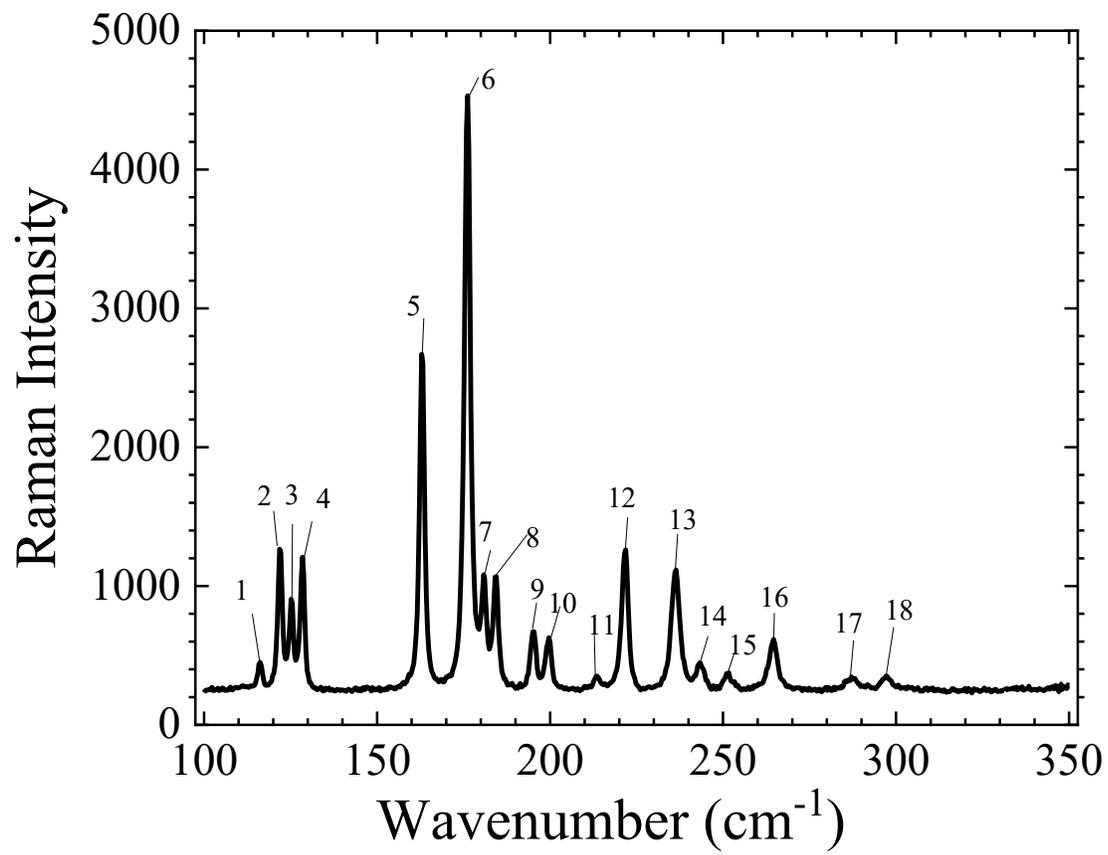

Fig. 5. Onasanya, *et al.*

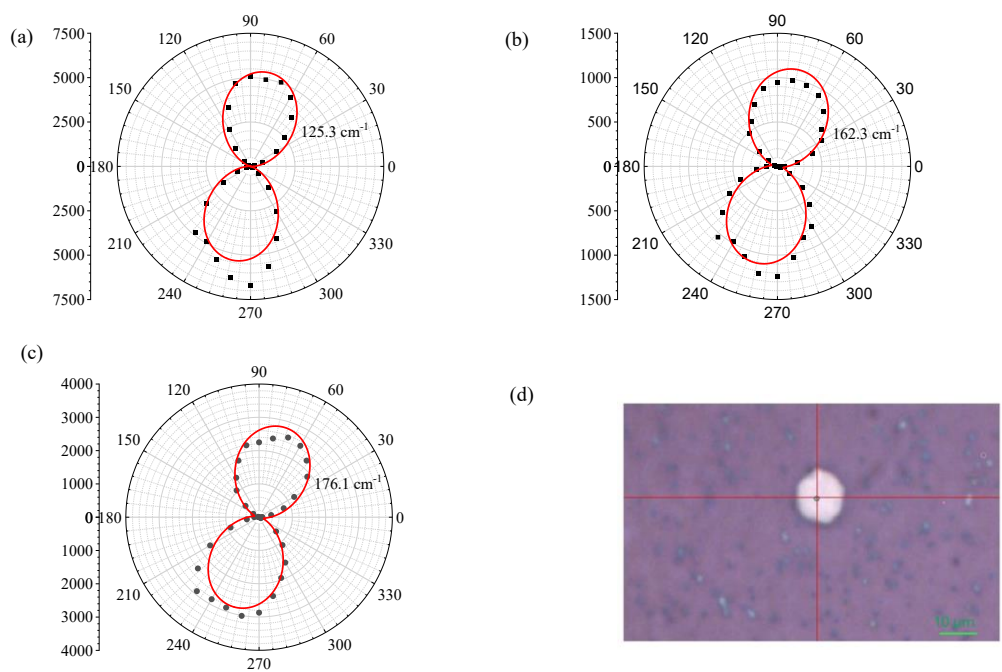

Fig. 6 Onasanya, *et al.*

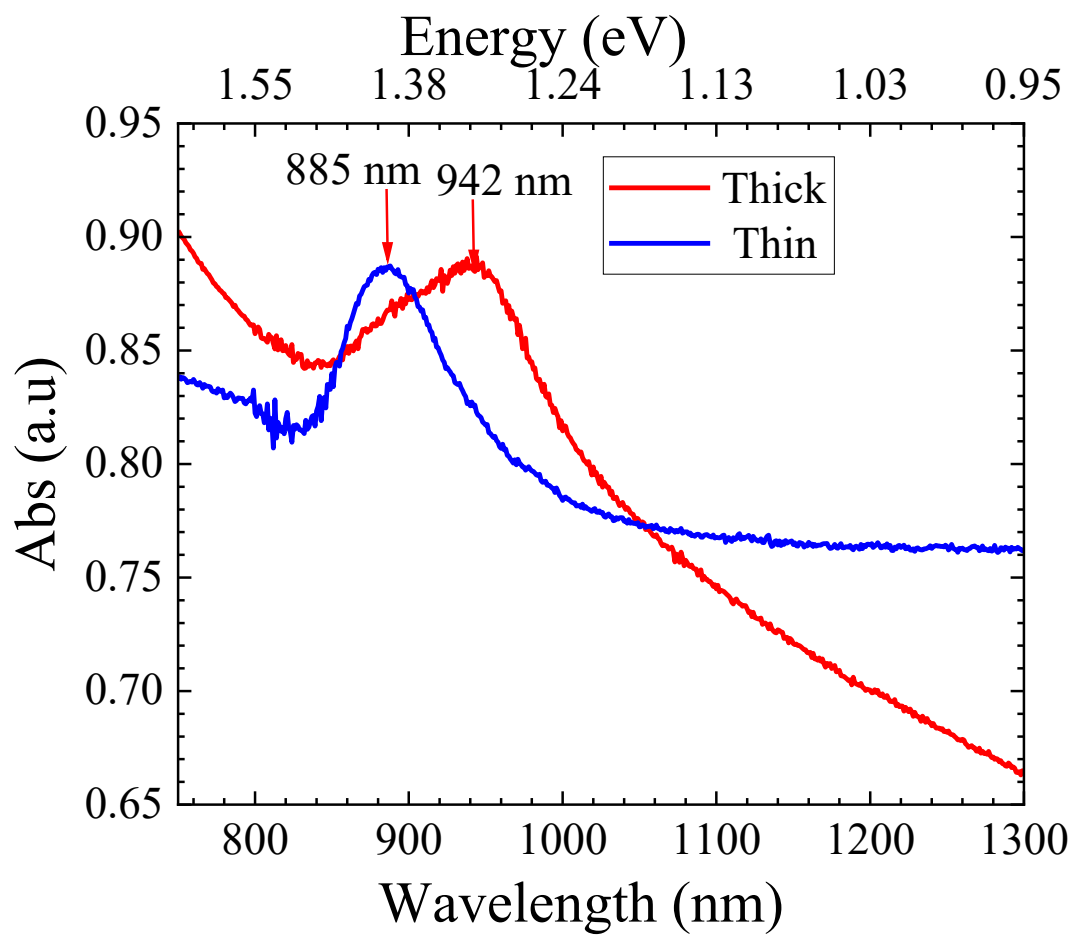

Fig. 7. Onasanya, *et al.*

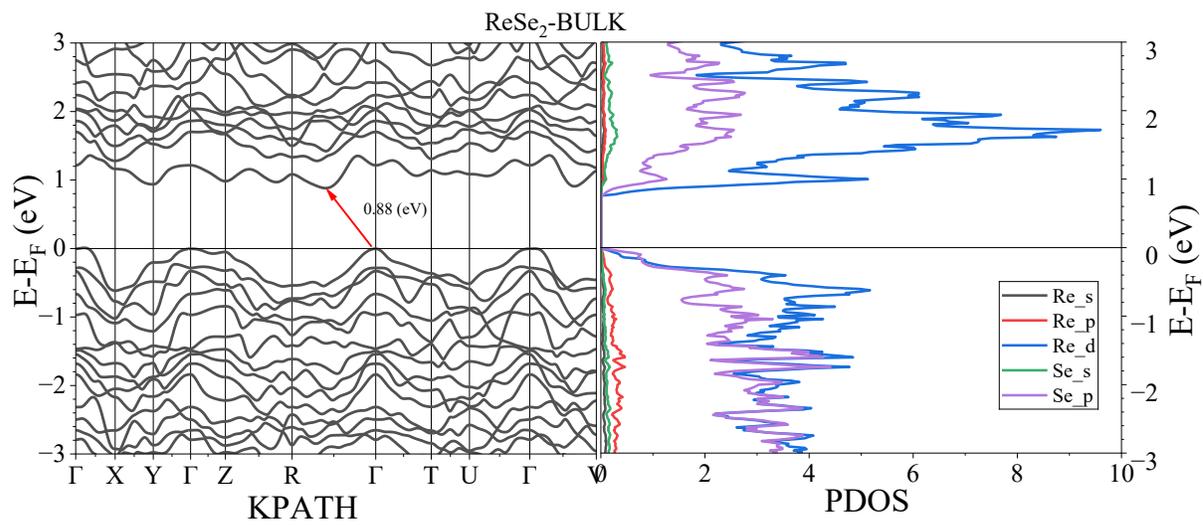

Fig. 8(a). Onasanya, *et al.*

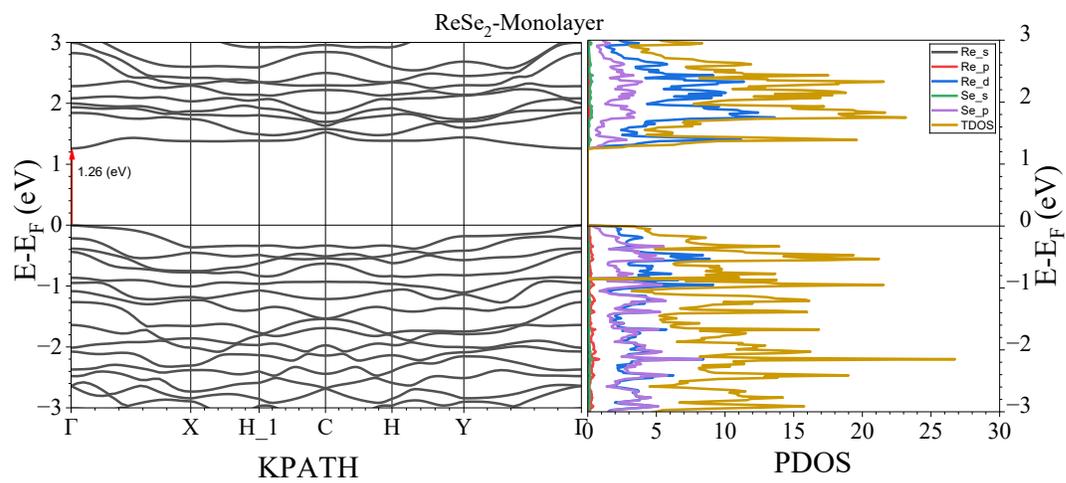

Fig. 8(b). Onasanya, *et al.*

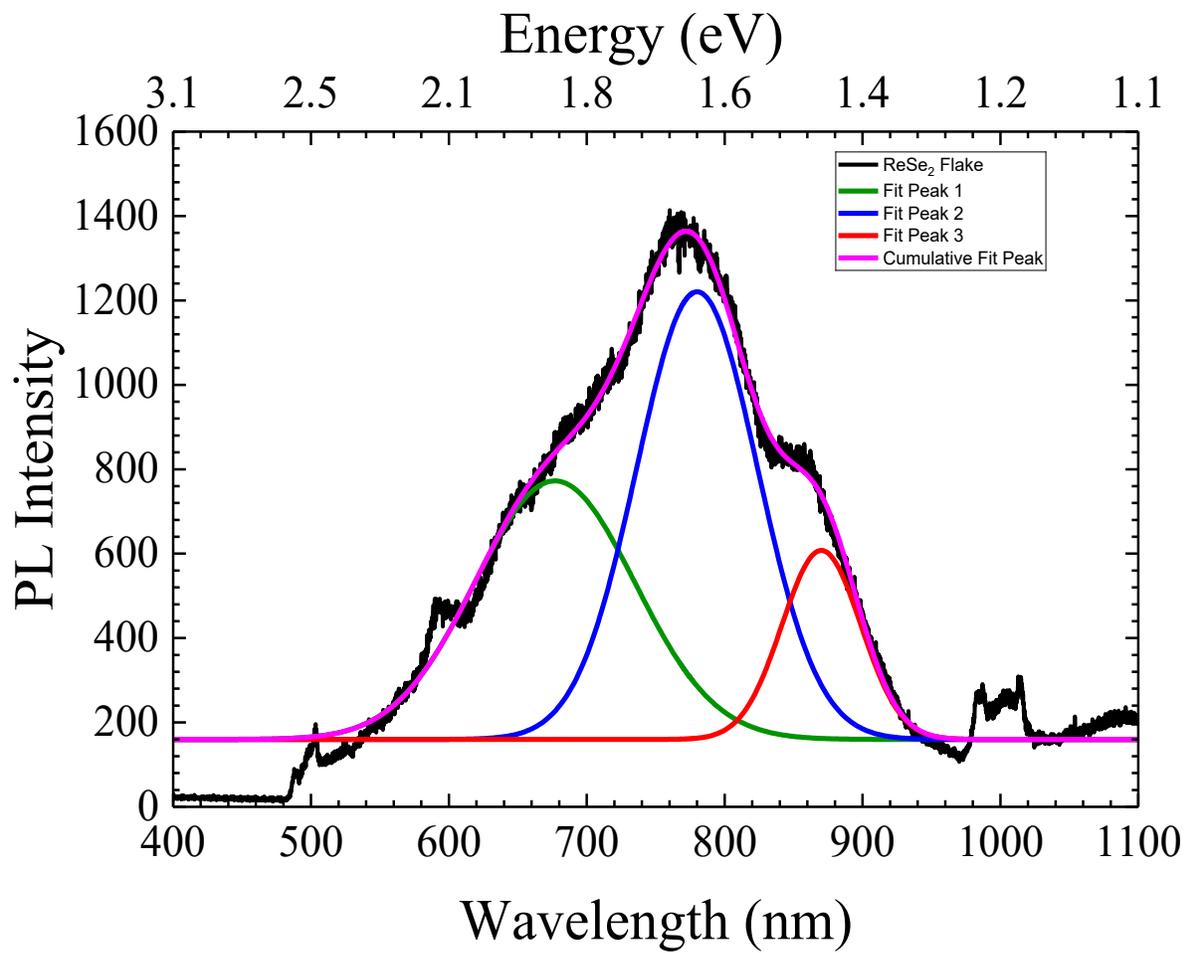

Fig. 9. Onasanya, *et al.*